\documentclass[aps,prl,preprint,superscriptaddress]{revtex4-1}

\usepackage{subeqn}
\usepackage{graphicx}

\newcommand{\bc}{{k_B}}

\newcommand{\Pe}{{\mbox{\upshape{Pe}}}} 
\newcommand{\tinyPe}{{\mbox{\scriptsize{\upshape{Pe}}}\:}} 
\newcommand{\pd}[2]{\frac{\partial #1}{\partial #2}}

\def\mathbi#1{\textbf{\em #1}}
\DeclareMathSymbol{\Delta}{\mathalpha}{letters}{"01}

\begin{document}


\title{Stochastic and Deterministic Vector Chromatography of Suspended Particles in 1D-Periodic Potentials}


\author{Jorge A. Bernate}
\email[]{jbernate@jhu.edu}
\affiliation{Department of Chemical and Biomolecular Engineering,Johns Hopkins University}

\author{German Drazer}
\email[]{drazer@jhu.edu}
\affiliation{Department of Chemical and Biomolecular Engineering,Johns Hopkins University}


\date{\today}

\begin{abstract}

We present a comprehensive description of vector chromatography that includes deterministic and stochastic transport in 1D-periodic free-energy landscapes, with both energetic and entropic contributions,
and highlights the parameters governing the deflection angle, i.e. the Peclet number and the partition ratio. 
We also investigate the dependence of the deflection angle on the shape of the free-energy landscape by varying the width of the  linear transitions in an otherwise dichotomous potential.
Finally, we present experimental results obtained in a microfluidic system in which gravity drives the suspended particles and, in combination with a bottom 
surface patterned with shallow rectangular grooves, creates a periodic landscape of (potential) energy barriers. 
The experiments validate the model and demonstrate that a simple, passive microdevice can lead to vector 
separation of colloidal particles based on both size and density.

\end{abstract}


\maketitle

Microfluidic systems for chemical and biological separation have shown great promise and opened the door for exciting new technologies. 
A number of separation systems based on driving suspended particles through 
a periodic stationary phase, for example, take advantage of the unprecedented 
control on the geometry and chemistry provided by micro-fabrication techniques. 
In fact, one-dimensional (1D) separation along the driving direction has been demonstrated 
in different microdevices with periodic stationary phase, ranging from entropic trap arrays \cite{Han:2000} 
to asymmetric structures acting as a ratchet \cite{Rousselet:1994}. 
The description of 1D transport of suspended particles past periodic entropy barriers, and to a lesser extent energy barriers, has also received 
considerable attention and rigorous results are available for the effective mobility of single particles
\cite{Zwanzig:1992,Reguera:2001,Reguera:2006,Yariv:2007,Wang:2009,Dorfman:2010,Pelton:2004,Ladavac:2004,Wang:2010,Burada:2010,Bernate:2011}.
Two-dimensional (2D) separation methods,  in which the different species constituting a sample migrate in different 
directions,  enabling their continuous fractionation and, in general, providing greater selectivity than 1D techniques,
have also been developed based on periodic stationary media, and have been categorized as vector chromatography (VC)~\cite{Dorfman:2001}. 
Notably, VC can be obtained in planar microfluidic devices via a straightforward extension of the aformentioned 1D methods,  
by driving the particles at an oblique angle with respect to the periodic direction.
This is the case when suspended particles are driven through force fields that are periodic in one of the directions of 
the separation plane and invariant in the other~\cite{Kralj:2006,Inglis:2004,Liu:2009}. 
Although a case-by-case analysis in the deterministic limit provided good agreement with experiments in these systems,
a general description is lacking.
In this letter, we extend previous results to obtain a comprehensive description of planar VC
in terms of the 1D-periodic free-energy of the system, including energetic and entropic contributions,
that captures the deterministic and Brownian limits. 
%
%
This unified description highlights the key parameters governing the migration angle of different species and
their relevance to the design and optimization of fractionation devices. 
We also present experimental results obtained in a  microfluidic system in which gravity not only 
drives the particles but also, in combination with a bottom surface patterned with shallow rectangular grooves, creates a periodic landscape of 
potential energy barriers.  The experiments agree well with the theoretical results, exhibit several of the qualitative features predicted by the model, and show the separation capability of the device.


Consider the transport of non-interacting Brownian particles in a planar microfluidic device, driven by a constant external force ${\mathbi{F}}$ (oriented at an angle $\theta_F=\arctan(F_y/F_x)$)  and moving through a potential energy landscape $V(x,y)$ that is periodic in one of the directions of the separation plane,
say $x$ axis ({\em periodic} direction), and invariant along the other, say $y$ axis ({\em uniform} direction). 
Let us assume that the driving force is contained in the separation plane, with any vertical component conveniently incorporated into the
potential $V$. 
The asymptotic distribution of particles in a unit cell is given by the steady-state solution of the Smoluchowski equation for
the {\em reduced} probability density $P^{\infty}({\mathbi{x}})$ \cite{MacroTransport,Li:2007},
\begin{equation}
0 = \nabla \cdot {\mathbi{J}}({\mathbi{x}}) = \nabla \cdot \left( {\mathbi{U}(\mathbi{x})} P^{\infty}({\mathbi{x}}) - 
{\mathbi{D}(\mathbi{x})}\cdot \nabla  P^{\infty}({\mathbi{x}}) \right),
\label{Eq:Smoluchowski}
\end{equation}
where ${\mathbi{J}}({\mathbi{x}})$ is the probability density flux, $\mathbi{U}({\mathbi{x}})$ is the instantaneous particle velocity 
and ${\mathbi{D}}({\mathbi{x}})$ is the diffusion tensor. In the low Reynolds number limit the velocity of the particle is a linear combination 
of the forces acting on it,  $\mathbi{U}({\mathbi{x}}) = \mathbi{M}(\mathbi{x})  \cdot \left( \mathbi{F} - \nabla V(\mathbi{x}) \right)$, 
where $\mathbi{M}(\mathbi{x})$ is the
mobility tensor, which locally satisfies the Stokes-Einstein relation $\mathbi{D}(\mathbi{x})= \bc T \mathbi{M}(\mathbi{x})$. 
$P^{\infty}({\mathbi{x}})$ is periodic in $x$, satisfies the no-flux condition in $z$, and is normalized, $\int_{\tau} P^{\infty}({\mathbi{x}}) dV = 1$, 
where $\tau$ is the volume of the unit cell. 
Given $P^{\infty}({\mathbi{x}})$ it is straightforward to compute the components of the average migration velocity  
$\overline{U}_{x,y} = \int_\tau J_{x,y} \; d\tau$ and the migration angle $\theta=\arctan(\overline{U}_y/\overline{U}_x)$, which is the relevant parameter in VC,
by means of macrotransport theory~\cite{MacroTransport}.

In planar microfluidic devices the particles are usually highly 
confined in the vertical direction, either geometrically, as in the case of entropic trapping, or due to particle-wall interaction potentials with narrow secondary minima \cite{Wang:2009,Wang:2010,Bernate:2011}. 
In this case, it is valid to assume fast equilibrium in the cross-section, which in this context is known as the
Fick-Jacobs (FJ) approximation \cite{DiffusionProcesses,Zwanzig:1992,Burada:2007,Laachi:2007} and is analogous to 
other projection methods that eliminate fast degrees of freedom 
\cite{FokkerPlanck,Nonequilibrium,Kalinay:2005}. 
It is then possible to write the probability distribution in terms of the marginal probability density,
${\cal P}(x)$, and the equilibrium conditional distribution in the cross-section, $\rho(z|x)_{eq}$, 
\begin{equation}
P^{\infty}(\mathbi{x}) \approx {\cal{P}}(x) \rho(z|x)_{eq} =  {\cal{P}}(x)  {\cal Q}^{-1} e^{-\beta V(x,z)},
\label{Eq:FickJacobsDensity}
\end{equation}
where $\beta=(\bc T)^{-1}$ and ${\cal Q}(x) = \int \exp(-\beta V(x,z))dz dy$ is the local partition function. 
In this approximation, the average migration velocity in the uniform direction is determined by the average mobility, 
that is $\overline{U}_y = \left(\int_0^{\ell_{x}} \left\langle M^{yy}\right\rangle_{eq} {\cal{P}}(x) dx \right) F_y$,  where 
$M^{yy}$ is the hydrodynamic mobility and, for any function $f(x,z)$, 
$\langle f \rangle_{eq} = \int f(x,z) \rho(z|x)_{eq} dz dy$ is the local equilibrium average over the cross-section.
Before we calculate the average velocity in the periodic direction, we note that $\overline{U}_x = \int^{\ell_x}_0 \mathcal{J}_x dx = \ell_x \mathcal{J}_x$, 
where $\mathcal{J}_x = \int\int J_x dydz$ is the total flux through any cross-section in the periodic direction and is constant in steady state.
Then, integrating Eq.~(\ref{Eq:Smoluchowski}) over the cross-section and using the FJ approximation we obtain
\begin{equation}
\mathcal{J}_x  = \left\langle M^{xx}\right\rangle_{eq}  \left\{ \left[ F_x - {\mathcal F}(x) \right]{\cal{P}} - \bc T\frac{d{\cal{P}}}{dx}  \right\},
\label{Eq:marginalJ}
\end{equation}
where ${\mathcal F}(x)$ is the mean force due to the potential, and following Zwanzig's derivation for the purely diffusive case~\cite{Zwanzig:1992},
we write it in terms of the local free-energy of the system, 
${\cal A}(x)=-\bc T \ln {\cal Q}(x)$,
\begin{equation}
{\mathcal F}(x) = -\left\langle \pd{V}{x}\right\rangle_{eq} = -\frac{\partial {\cal A}(x)}{\partial x}.
\label{Eq:MF}
\end{equation}
%

It is clear from Eq.~(\ref{Eq:marginalJ}) that the total flux in the periodic direction, and therefore $\overline{U}_x$, 
have both diffusive and convective contributions.
The first convective term shows that, as expected, even in the absence of a periodic potential (i. e. ${\mathcal F}\equiv0$), 
a macroscopic anisotropy in the mobilities, $\left\langle M^{xx}\right\rangle_{eq} \neq \left\langle M^{yy}\right\rangle_{eq}$,
could lead to a non-zero deflection angle, $\Delta\theta=\theta-\theta_F\neq0$.
Furthermore, the dependence of the average mobility on $x$ can also result in a 
non-zero deflection angle, independent of the local isotropy of the mobility tensor, given that the contribution of the diffusive flux to $\overline{U}_x$ (last term in Eq.~(\ref{Eq:marginalJ})) would not vanish in this case~\cite{Bernate:2011}.
For simplicity, however, we shall assume that the mobility functions are constant and equal, $M^{xx} = M^{yy}= M$. In this case,
we have $\overline{U}_y = M F_y$ and it is clear that a non-zero deflection angle implies that $\overline{U}_x \neq M F_x$, and vice versa. 
Moreover, in this case the diffusive contribution to $\overline{U}_x$ vanishes, and therefore only ${\mathcal F}(x)$ can contribute to a non-zero 
deflection angle. Solving Eq.(\ref{Eq:marginalJ}) \cite{FokkerPlanck,Bernate:2011} we obtain,
\begin{equation}
\tan \theta \!=\! \tan \theta_F \! 
\left[ \frac{\Pe}{1-e^{-\tinyPe}}   
\! \! \int_0^{1}\!\!\!   d\tilde{x} \, e^{-\beta {\mathcal{A}}'\!(\tilde{x})} 
\int_{\tilde{x}}^{\tilde{x}+1} \!\!\!\! \!\!\!\! \! d\xi \, e^{\beta {\mathcal{A}}'\!(\xi)} \right] \! ,\!
\end{equation}
where  $\beta{\mathcal A}'\!(\tilde{x}) = \beta \mathcal{A}(\tilde{x}) - \Pe \tilde{x}$, and $\tilde{x}=x/\ell_x$.
The P\'eclet number, $\Pe=\beta F_x \ell_x$, measures the relative magnitude of convective to
thermal transport in the system and is one of the two dimensionless parameters dictating the behavior of the migration angle.
Alternatively, one can use a characteristic value of the mean force, such as its maximum value ${\cal{F}}_{max}$, 
and consider the normalized driving force $f =F_x/{\cal{F}}_{max}$ as an
independent parameter. However, we shall see below that the P\'eclet number and the normalized force are complementary 
for the description of a given system, in that one is the appropriate parameter to consider when the other diverges. 
The second parameter is the partition ratio $\mathcal{K}$, which measures the spatial variations in the distribution of particles in equilibrium. 
Its importance becomes apparent when we adopt the amplitude of variations in the free-energy, $\Delta \cal{A}$, 
as the characteristic energy scale to write $\beta{\mathcal A}(\tilde{x}) = (\ln {\mathcal{K}}) \, \tilde{\mathcal{A}}(\tilde{x})$, 
with ${\mathcal{K}} = \exp(\beta \Delta {\mathcal{A}})$, which in the context of transition-state theory, corresponds to an Arrhenius factor~\cite{Hanggi:1990}. 

\begin{figure}
\includegraphics[width=\columnwidth]{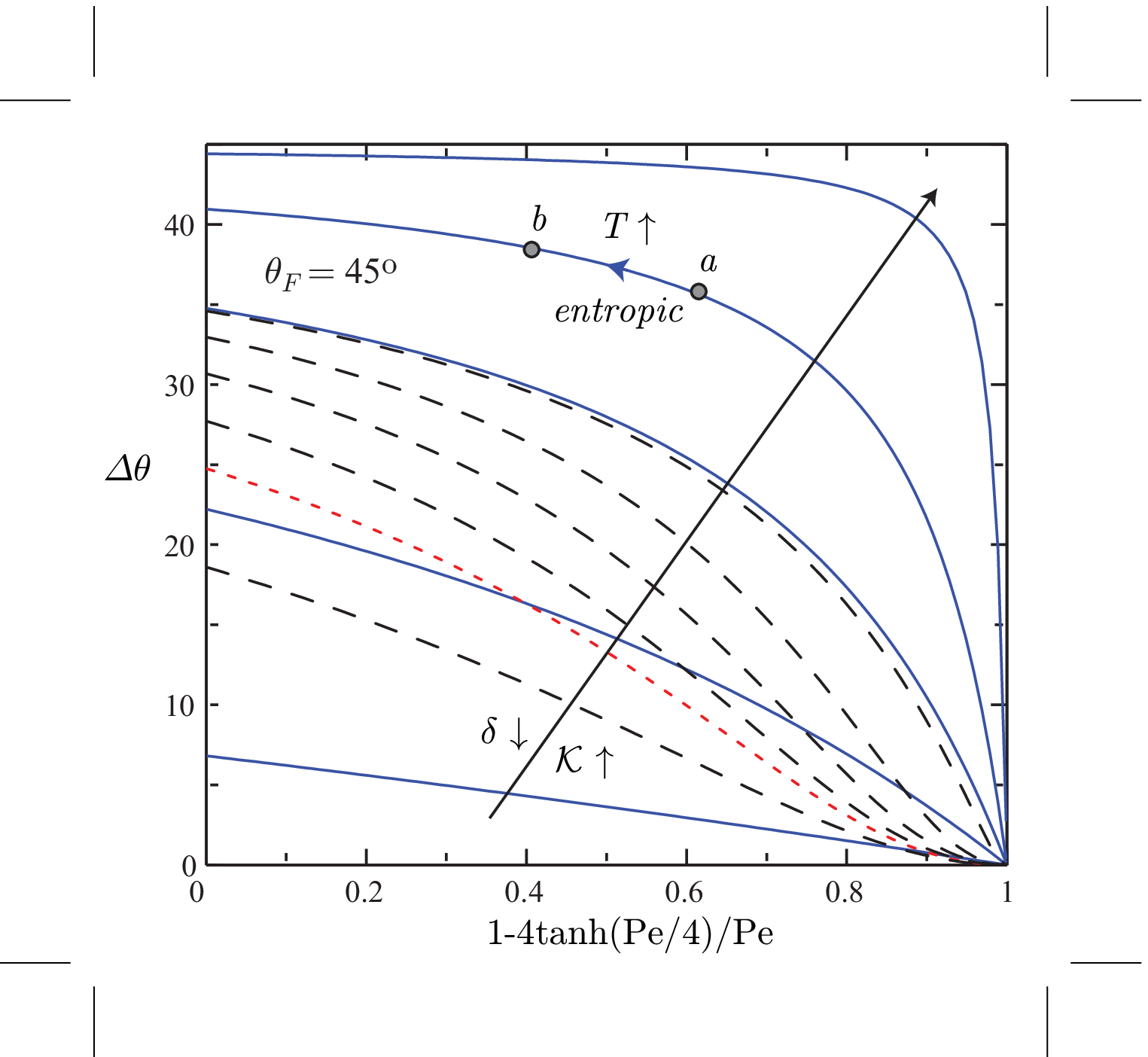}
\caption{\label{fig:Dtheta_Pe} Deflection angle as a function of the P\'eclet number. Solid lines correspond to the SW potential 
with $\log {\cal{K}} = \beta \Delta {\mathcal{A}} = 1, 2, 3, 4$ and $6$. The dashed lines correspond to the LTD potential with $\beta \Delta {\mathcal{A}} = 3$, transition regions $\delta = 0.01, 0.1, 0.2, 0.3, 0.5$, and $\epsilon_1 = \epsilon_2 = 0.5-\delta$. The arrow traverses curves of increasing ${\cal{K}}$ for the SW potential and curves of decreasing $\delta$ for the LTD potential. The evolution ($a \to b$) of a purely entropic system upon a temperature increase is shown.}
\end{figure}

In order to investigate the role of these parameters on the migration angle
we consider a {\em cosine} potential, 
$\tilde{\mathcal A}(\tilde{x})= 1/2\cos 2\pi \tilde x$, and a {\emph{dichotomous potential with linear transitions} (LTD potential), given by regions of constant potential, $\tilde{\mathcal A}=0$ and $\tilde{\mathcal A}= 1$, connected by linear transitions of width $\delta$ (see inset in Fig.~\ref{fig:Dtheta_f}). The LTD potential in the limit $\delta=0$ corresponds to a {\em square wave} (SW) potential. 
In all cases, the effect of the periodic potential is to reduce the average velocity in the periodic direction~\cite{Reimann:2002}, $\overline{U}_x$, 
resulting in positive deflection angles (for the cases shown here in which $\theta_F=45^\circ$, this implies that $0^\circ < \Delta \theta < 45^\circ$).
Fig.~\ref{fig:Dtheta_Pe} shows the deflection angle as a function of $\Pe$ for the SW (solid curves), LTD (dashed lines), and cosine (dotted line) potentials. It is clear that the deflection angle decreases with $\Pe$ and increases with $\cal{K}$ (arrow direction), independent of its entropic or energetic origin. 
Fig.~\ref{fig:Dtheta_Pe} also shows that for a given ${\cal{K}}$ and $\Pe$, the smaller the transition region in the LTD potential
the higher the deflection angle, with, as expected, $\Delta \theta$ converging to the curve for the SW potential in the limit $\delta \to 0$ (arrow direction).  
In Fig.~\ref{fig:Dtheta_f} we plot the deflection angle as a function of the normalized force and consider the effect of Brownian motion 
for the different potentials. Specifically, we compare the deflection angle obtained at a finite partition ratio for the cosine and LTD potentials 
with that in the deterministic limit ($\Pe \to \infty$ and finite $f$). Note that Figs.~\ref{fig:Dtheta_Pe} and~\ref{fig:Dtheta_f} are complementary, 
in that they allow us to investigate independent limits, i. e., the SW potential limit($f\to0$ and finite $\Pe$) and the deterministic limit ($\Pe\to\infty$ and finite $f$), respectively. 
In all cases, the deflection angle decreases with $f$, which is similar to the behavior observed with $\Pe$ in Fig.~\ref{fig:Dtheta_Pe}. Fig.~\ref{fig:Dtheta_f} also shows that Brownian motion allows the particles to cross the potential barriers even when $f\leq1$. This is in contrast to the deterministic case, in which particles are locked to move along the uniform direction. We also investigate the effect that the transition region $\delta$ has on the deflection angle. In the deterministic limit, the deflection angle for the LTD potential has the simple analytical expression~\cite{EPAPS},
\begin{equation}
\frac{\tan \theta}{\tan\theta_F } \!=\!  \left[ 1\!-\! 2\delta \!+\! \frac{f \delta}{f\!-\!1} \!+\! \frac{f \delta}{f\!+\!1} \right] \! 
= \! \left[ 1\!+\! \frac{2 \delta}{f^2 \!-\!1} \right] ,
\label{transit}
\end{equation}
and it is clear that larger transition regions lead to larger deflection angles for any $f>1$. 
The reason is that as $\delta$ increases (at constant $f$), the particle is deflected for a longer time as it crosses the potential barrier.
Note the difference in the contributions coming from the regions with $\pm f$ to the transit time, 
as shown by the corresponding terms in Eq.~\ref{transit}.
In the presence of Brownian motion we observe the same trend for large driving forces $f$, as expected. 
On the other hand, as the driving force decreases and barrier hoping is dominated by thermal motion, the behavior reverses
and larger transition regions lead to smaller deflection angles. This crossover between the deterministic and Brownian
 cases as $f$ decreases is consistent with the behavior observed as a function of $\Pe$ in Fig. ~\ref{fig:Dtheta_Pe}. In fact, the limits
 $\Pe \ll 1$ and $f \ll 1$ coincide in the linear response regime, where the reduction in mobility (and diffusivity)
 is given by $\left( \int {\cal Q} dx \int {\cal Q}^{-1} dx \right)$~\cite{Parris:1997}.
\begin{figure}
\includegraphics[width=\columnwidth]{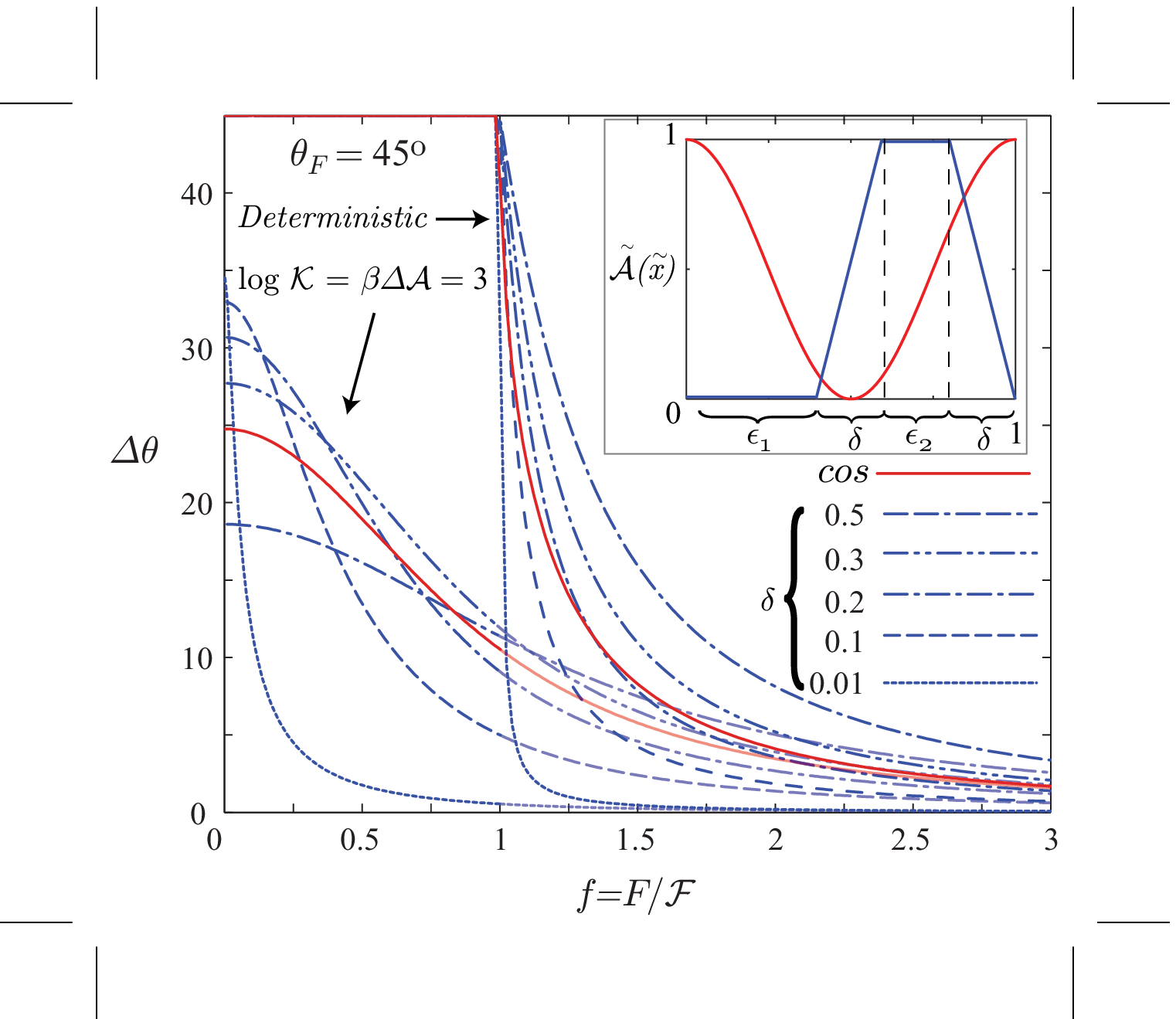}
\caption{\label{fig:Dtheta_f} Effect of $\delta$ on the deflection angle as function of the normalized force in the stochastic and deterministic regimes.  The solid lines correspond to the cosine potential while the dashed lines correspond to the LTD potential for different transition regions and $\epsilon_1 = \epsilon_2 = 0.5-\delta$. The curves corresponding to the Brownian case and the respective deterministic curves have the same line style. The inset shows schematics of the cosine and the LTD potentials.}
\end{figure}
Fig.~\ref{fig:Dtheta_f_delta_0p1} shows the effect of the partition ratio on the deflection angle for the LTD potential with a given $\delta$. 
Clearly, the deflection angle increases as the partition ratio increases, converging to an asymptotic curve for ${\cal{K}}\to\infty$. 
This upper limit coincides with the deterministic limit for a purely energetic potential of mean force~\cite{Pelton:2004}. 
In terms of separation devices, it becomes clear that in order to obtain large deflection angles and high selectivity 
 it is desirable to operate around $f\lesssim 1$. 
 
 The results presented in Figs.~\ref{fig:Dtheta_Pe}--\ref{fig:Dtheta_f_delta_0p1} also reveal the role of temperature
 in different separation systems~\cite{Reguera:2006}. In entropic trapping ($\Delta {\mathcal{A}} \propto k T$), for example, the partition ratio is determined by the ratio between the available configurations in the slit and well regions~\cite{Dorfman:2010}, and is thus independent of temperature. On the other hand, both $\Pe$ and $f$ decrease with temperature,
thus leading to larger deflection angles. In fact, increasing the temperature for a given geometry of the entropic traps
corresponds to the system moving along curves of constant partition ratio, as shown in Figs.~\ref{fig:Dtheta_Pe} and ~\ref{fig:Dtheta_f_delta_0p1}. 
On the other hand, in the purely energetic case $\Delta {\mathcal{A}}$ and $f$ are independent of temperature.
Therefore, for a given potential energy landscape, increasing the temperature reduces the partition ratio and the deflection angle. 
This corresponds to the system moving along curves of constant $f$ and decreasing partition ratio, that is vertical lines in Fig.~\ref{fig:Dtheta_f_delta_0p1}, as shown. 

\begin{figure}
\includegraphics[width=\columnwidth]{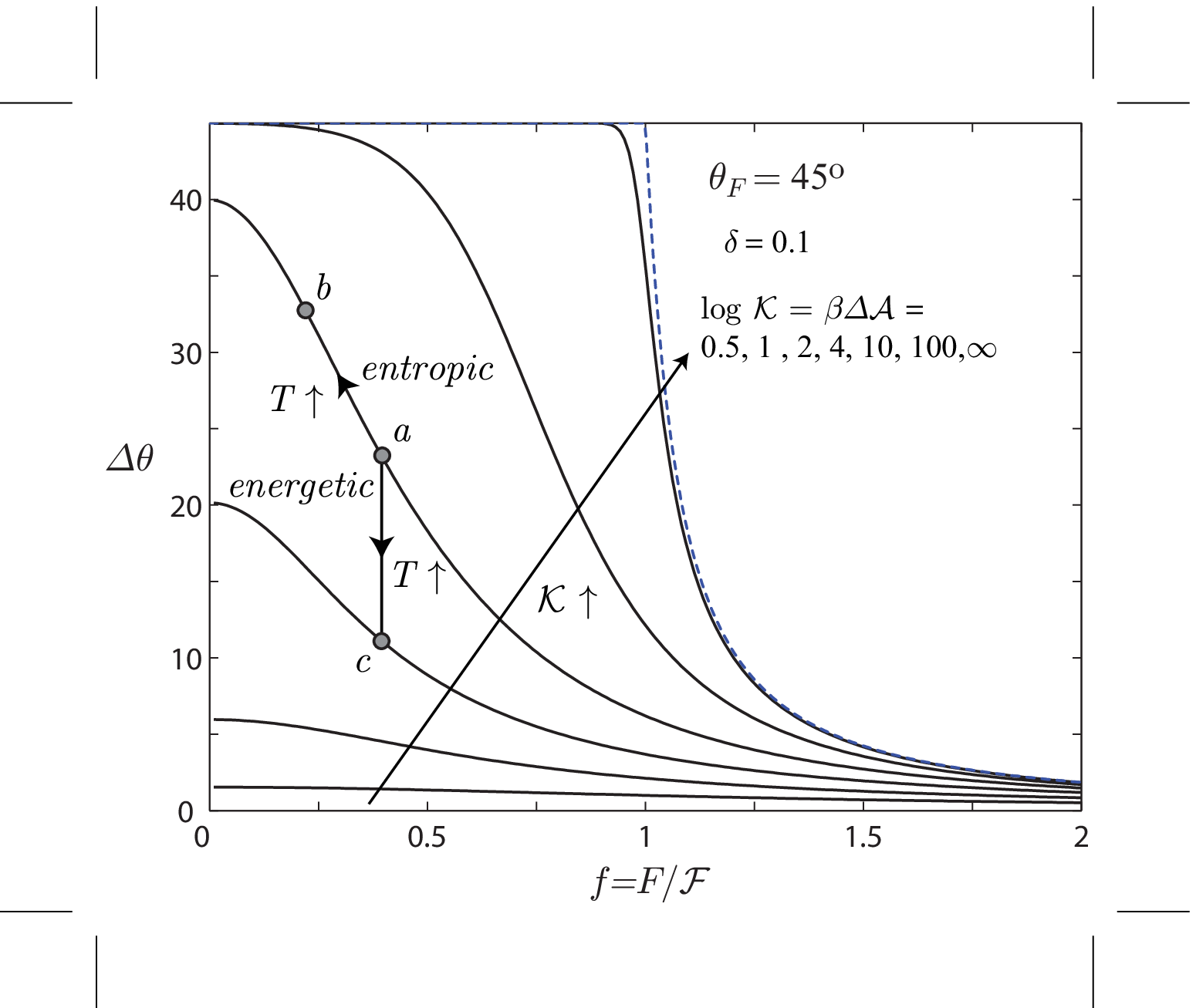}
\caption{ Effect of the partition coefficient on the deflection angle as a function of the normalized force for the LTD potential for a given width of the transition region and $\epsilon_1 = \epsilon_2 = 0.5-\delta$. The arrow traverses curves of increasing partition ratio. The dashed curve corresponds to the deterministic limit. Upon increasing the temperature, a system at point $a$ ``moves'' along the same curve toward point $b$ for a purely entropic potential, and vertically down toward point $c$ for a purely energetic potential.}
\label{fig:Dtheta_f_delta_0p1}
\end{figure}

%

\begin{figure}
\includegraphics[width=\columnwidth]{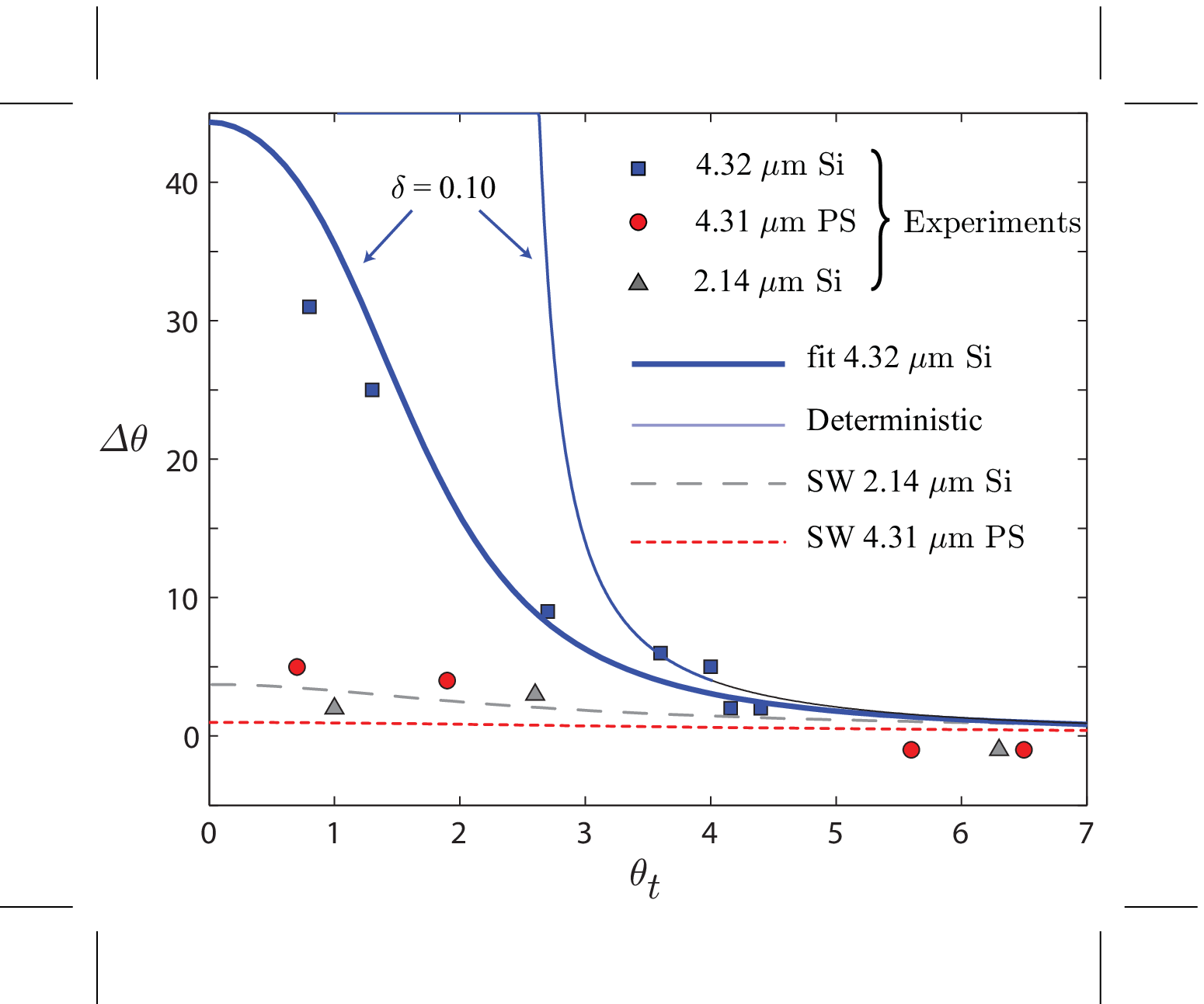}
\caption{\label{fig:Experiments} Deflection angle as a function of the tilt angle. 
The circular (dashed)  and triangular (dotted) symbols (curves)  correspond to the experimental results (SW potential) for the 4.31 $\mu$m polystyrene particles and to the 2.14 $\mu$m silica particles, respectively. The solid dark line is a fit of the experimental data to the results of the LTD potential for the 4.32 $\mu$m silica particles  (square symbols) using the width of the transition region as a fitting parameter ($\delta = 0.10$), while the light solid line corresponds to the respective deterministic limit. See supplementary information for the standard deviations and for a video showing a representative experiment for the 4.32 $\mu$m silica particles.}
\end{figure}

We performed experiments in a microfluidic system in which suspended particles are driven over a glass 
surface on which we had (isotropically) etched rectangular grooves in parallel. The periodicity is 20 $\mu$m, 
the shallow grooves are $\Delta{\cal H}=65$ nm deep and  13 $\mu$m wide. 
The channel is high enough to neglect confinement effects. Gravity induces periodic energy barriers due to the presence of grooves, and also drives the particles by tilting the bottom surface an angle $\theta_t$. 
We use silica particles (4.32 $\mu$m and 2.14 $\mu$m diameter) and polystyrene particles (4.31 $\mu$m diameter). 
The gravity-induced partition ratio for a particle of radius $a$ is given by ${\cal{K}} = \exp \left( 4/3 \pi a^3 \Delta\rho \: g \cos\theta_t \: \Delta {\cal{H}}/\bc T\right)$ 
where $\Delta \rho$ is the buoyant density of the particles and $g$ is the acceleration due to gravity~\cite{Bernate:2011}.  
For $\theta_t  = 0^{\circ}$ the partition coefficient of the  4.32 $\mu$m silica particles is
at least two orders of magnitude larger than that for the 2.14 $\mu$m silica and 4.31 $\mu$m polystyrene particles. 
Thus, at small tilt angles, the 4.32 $\mu$m silica particles should experience much larger deflections than the other particles,
which would demonstrate that it is possible to fractionate particles by size or density.
In Fig.~\ref{fig:Experiments} we show the experimental results for the deflection angle,for a forcing angle $\theta_F = 45^{\circ}$, as a function of the tilt angle (note that unlike $\Pe$ and $f$, the tilt angle is common to all particles in a given experiment).
The theoretical curve corresponds to the LTD potential, 
with $\delta$ calculated from the best fit to the experimental data and representing an effective transition 
region in the interaction between a suspended particle and the bottom grooves.
We obtain good agreement for the 4.32 $\mu$m silica particles
with $\delta = 0.10 \pm 0.01$ (2.0 $\pm$ 0.2 $\mu$m), which compares well with an order of magnitude estimate $\delta_{zoi} = 2\sqrt{2 (a+ h_{e})\left(\Delta {\cal{H}} + \kappa^{-1}\right)} = 1.4$ $\mu$m obtained by extending the concept of a {\emph{zone of influence}}~\cite{Kozlova:2006} for a particle suspended at its equilibrium separation from the wall ($h_{e} = 259$ nm) and in the vicinity of a step~\cite{EPAPS}.
The theoretical curves for 2.14 $\mu$m silica and 4.31 $\mu$m polystyrene particles 
are insensitive to the width of the transition region, with differences smaller than 2.5$^\circ$,
and the data is compared to SW potentials, with good agreement.
In these latter cases, the particles easily overcome the energy barriers due to thermal fluctuations, significantly reducing confinement effects and leading to small deflection angles. These experiments show that gravity, along with a bottom surface patterned with slanted periodic grooves, could be used to separate particles according to their mass. More generally, the role of gravity can be effected or enhanced by other fields, e.g. electric, dielectrophoretic, magnetic, etc., potentially leading to a versatile separation technique.

\begin{acknowledgments}
The authors would like to acknowledge Prof. Dilip Asthagiri for helpful discussions. This material is based upon work supported by the National Science Foundation under grants No. CBET-0954840 and CBET-0731032.
\end{acknowledgments}

%

\end{document}